\documentclass[12pt]{article}
\usepackage[margin=1 in]{geometry}  % set the margins to 1in on all sides
\usepackage{graphicx}              % to include figures
\usepackage{amsmath,bigints}               % great math stuff
\usepackage{amsfonts}              % for blackboard bold, etc
\usepackage{amsthm}                % better theorem environments
\usepackage{caption,fancyhdr} 
\usepackage{multirow,hyperref}
\usepackage{lineno,hyperref}
\usepackage{authblk}
\usepackage[justification=centering]{caption}

\newtheorem{thm}{Theorem}
\newtheorem{lem}{Lemma}

\newtheorem{defn}{Definition}
\newtheorem{assmp}{Assumption}

\title{\textbf{Ratio of two independent xgamma random
variables and some distributional properties }}
\author[a]{Subhradev Sen\thanks{E-mail: subhradev.stat@gmail.com}}
\author[b]{Suman K. Ghosh\thanks{E-mail: suman.ghosh.2006@gmail.com}}
\author[c]{Hazem Al-Mofleh\thanks{E-mail:halmofleh@eureka.edu}}
\affil[a,b]{\small{Alliance School of Applied Mathematics, Alliance University, Bengaluru, India.}}
\affil[c]{\small{Division of Science and Mathematics, Eureka College, IL, USA.}}
\date{}
\begin{document}
\maketitle

\begin{abstract}

In this investigation, the distribution of the ratio of two independently distributed xgamma (Sen et al.~2016) random variables X and Y , with different parameters, is proposed and studied. The related distributional properties such as, moments, entropy measures, are investigated. We have also shown a unique characterization of the proposed distribution based on truncated incomplete moments.
\end{abstract}

\clearpage
\section{Introduction}

The distribution of the ratio of random variables is of interest to problems in the biological and physical sciences, econometrics, classification, ranking, and selection. Examples of the use of the ratio of random variables are Mendelian inheritance relationships in genetics, mass-energy relationships in nuclear physics, the purpose of rainfall in meteorology, and stock relationships in economics, see Ali et al.~(2007), Bowman et al.~(1998) and the references therein. In engineering applications, ratio of two random variables can be found easily. For example, a very prominent example of ratio of random variables is the stress-strength model in the context of reliability engineering.\\
It describes the life of a component which has a random strength $Y$ and is subjected to random stress $X$. The component fails at the instant that the stress applied to it exceeds the strength and the component will function satisfactorily as long as $Y>X$. In such situation, $Pr(X<Y)$ provides a measure of component reliability. Stress-strength reliability has several applications especially in engineering concepts, such as structures, deterioration of rocket motors, fatigue (static) of ceramic components, fatigue failure in aircraft structures, and the ageing of concrete pressure vessels, to mention a few. 
As other examples, ratios are often used to denote relative performance of a desired quality or as a value often specified in a design trade-off. Two common examples are, (i) the \textit{aspect ratio}, the ratio of its length to its width, and (ii) the \textit{performance ratio}, ratio of two variables measuring the performance of two different methods for one given job. For more description on these ratios see Nadarajah~(2010), Sackey and Smith (2009), Worzyk et al.~(1997) and references therein.\\
Among many other lifetime model, xgamma distribution (Sen et al.,~2016) is a well established probability distribution that has flexibility in modelling data sets coming from diverse areas of application. A notable amount of research works can be found on xgamma distribution, methodology, its extensions, and applicability (see for more insights, Sen et al., 2017, 2018, 2019; Cordeiro et al., 2020; Yadav et al., 2021; Abulebda et al., 2022; Alizadeh et al., 2023; Pathak et al.,~2023). We have the following definition for xgamma distribution with one parameter.
\begin{defn}
A continuous random variable $T$ is said to follow a xgamma distribution with parameter $\theta (>0)$ if its probability density function (pdf) is given by
\begin{align}
\label{pdfxgd}
f(t)=\frac{\theta ^{2}}{(1+\theta )}\left( 1+\frac{\theta }{2}{t^{2}}\right)
e^{-\theta t},\quad t>0.
\end{align}
\end{defn}
The corresponding cumulative density function (cdf) is given by
\begin{align}
F(t)=1-\frac{\left( 1+\theta +\theta t+\frac{\theta ^{2}t^{2}}{2}\right) }{%
(1+\theta )}e^{-\theta t},\quad t>0,\theta >0.
\end{align}
No attempt has been made to study the distribution of the ratio of two random variables following density function given in (\ref{pdfxgd}). Our objective in this investigation is to propose and study the distribution of ratio of two xgamma variables with different parameters. We assume that the two variables, viz., $X$ and $Y$, are independent. However, this assumption
may not always be realistic owing to the fact that, for large samples, the distribution assumption of independence is consistent with that of non-independence, as can be seen in Cox and Hinkley~(1979). 

The rest of the article is organized as follows. The synthesis of the distribution of the ratio of two independent xgamma variables is described and presented in the section~\ref{sec:2} and its dedicated subsections. Moments and related measures have been studied in section~\ref{sec:3}. Section~\ref{sec:4} deals with important entropy measures. Method of maximum likelihood is proposed in section~\ref{sec:5} for estimating the unknown parameters of the proposed distribution. Real life data sets have been analyzed to understand the applicability of the distribution in section~\ref{sec:6}. Finally, section~\ref{sec:7} concludes.

\section{Synthesis of the distribution for $Z=|\frac{X}{Y}|$}
\label{sec:2}
In this section we derive the distribution of the ratio of two independent xgamma variables with following specifications.\\
Let $X$ and $Y$ be two independent xgamma random variables with parameters $\alpha$ and $\beta$ respectively. The probability density functions of $X$ and $Y$ are given by  
\begin{eqnarray} \label{eqpdf1}
f_X(x)=\frac{\alpha^2}{1+\alpha}\left(1+\frac{\alpha}{2}x^2 \right)e^{-\alpha x}
\end{eqnarray} and
\begin{eqnarray} \label{eqpdf2}
f_Y(y)=\frac{\beta^2}{1+\beta}\left(1+\frac{\beta}{2}y^2 \right)e^{-\beta y}
\end{eqnarray}
respectively, for $x>0,y>0,\alpha >0$ and $\beta >0$.\\ We define $Z=|\frac{X}{Y}|$, the form of the ratio, as a new random variable which is non-negative.
\subsection{Distribution function of the ratio $Z=|\frac{X}{Y}|$}
\begin{thm} \label{Th1}
Suppose that $X$ and $Y$ are independent random variables distributed according to (\ref{eqpdf1}) and (\ref{eqpdf2}), respectively. The cumulative density function (cdf) of $Z$ can be expressed as
\begin{eqnarray} \label{eqcdf1}
F_Z(z)=1-\bar{F_Z}(z)
\end{eqnarray} where, 
\begin{align*}
\bar{F_Z}(z)=\left(\frac{\beta^2}{1+\beta} \right)\frac{1}{(\alpha z+\beta)}
\Biggl[ 1+\frac{\alpha z}{(1+\alpha)(\alpha z+\beta)}+\frac{\left(\frac{\alpha^2 z^2}{1+\alpha}+\beta \right)}{(\alpha z+\beta)^2}+\frac{3\alpha \beta z}{2(1+\alpha)(\alpha z+\beta)^3}\\+\frac{6\alpha^2\beta z^2}{(1+\alpha)(\alpha z+\beta)^4} \Biggr].
\end{align*}
\end{thm}
\begin{proof}
$Z=|\frac{X}{Y}|=\frac{X}{Y}$ since both the random variables are non-negative valued. Therefore,
\begin{eqnarray*}
F_Z(z)&=&P(Z \leq z)=P\left(\frac{X}{Y} \leq z\right)=\int_{0}^{\infty}{P(X \leq zY|Y=y)P(Y=y)dy}\\ 
&=&\int_{0}^{\infty} {P(X \leq zy)P(Y=y)dy = \int_{0}^{\infty}F_X(zy)f_Y(y)dy}\\ 
&=&\int_{0}^{\infty}\left\lbrace 1-\frac{1+\alpha+\alpha zy+\left(\frac{\alpha^2 z^2}{2}\right)y^2}{1+\alpha}e^{-\alpha zy} \right\rbrace f_Y(y)dy\\ 
&=&1- \int_{0}^{\infty} \frac{1+\alpha+\alpha zy+\left(\frac{\alpha^2 z^2}{2} \right)y^2}{1+\alpha}e^{-\alpha zy}f_Y(y)dy =1-I_1,
\end{eqnarray*}
where,
\begin{eqnarray*}
I_1 &=&\int_{0}^{\infty} \frac{1+\alpha+\alpha zy+\left(\frac{\alpha^2 z^2}{2} \right)y^2}{1+\alpha}e^{-\alpha zy} \frac{\beta^2}{1+\beta}\left(1+\frac{\beta}{2}y^2 \right)e^{-\beta y}dy, \\ 
&=&\frac{\beta^2}{1+\beta} \int_{0}^{\infty} \left\lbrace 1+\frac{\alpha z}{1+\alpha} \left(y+\frac{\alpha z}{2}y^2 \right) \right\rbrace \left(1+\frac{\beta}{2}y^2 \right)e^{-(\alpha z+\beta)y}dy \\ 
&=&\frac{\beta^2}{(1+\beta)(\alpha z+\beta)}I_2,
\end{eqnarray*} 
where,
\begin{align*}
I_2=\Gamma(1)+\frac{\Gamma(2)\alpha z}{(1+\alpha)(\alpha z+\beta)}+\frac{\Gamma(3)}{2}\frac{\left(\frac{\alpha^2 z^2}{1+\alpha}+\beta \right)}{(\alpha z+\beta)^2}+\frac{\Gamma(4)\alpha \beta z}{2(1+\alpha)(\alpha z+\beta)^3}+\frac{\Gamma(5)\alpha^2 \beta z^2}{4(1+\alpha)(\alpha z+\beta)^4}.
\end{align*}
On simplification, we obtain (\ref{eqcdf1}).
\end{proof}
\subsection{Probability density function of the ratio $Z=|\frac{X}{Y}|$}
\begin{thm} \label{Th2}
Suppose that $X$ and $Y$ are independent random variables distributed according to (\ref{eqpdf1}) and (\ref{eqpdf2}), respectively. The pdf of $Z$ can be expressed as 
\begin{eqnarray} \label{eqpdf3}
f_Z(z)=\left(\frac{\alpha^2}{1+\alpha} \right)\left(\frac{\beta^2}{1+\beta} \right)\left[\frac{1}{(\alpha z+\beta)^2}+\frac{3\left(\alpha z^2+\beta \right)}{(\alpha z+\beta)^4}+\frac{30\alpha \beta z^2}{(\alpha z+\beta)^6} \right] 
\end{eqnarray}
\end{thm}
\begin{proof}
Let $Z=|\frac{X}{Y}|, V=X$. Therefore, $f_{Z,V}(z,v)=|\frac{\partial (x,y)}{\partial (z,v)}|f_{X,Y}(x,y)=\frac{v}{z^2}f_X(x)f_Y(y)$. \\
The pdf of $Z$ is obtained as follows 
\begin{eqnarray*}
f_Z(z)&=&\int_{0}^{\infty} f_{Z,V}(z,v)dv = \int_{0}^{\infty} \frac{v}{z^2}f_X(x)f_Y(y)dv = \int_{0}^{\infty} \frac{v}{z^2}f_X(v)f_Y\left(\frac{v}{z}\right)dv\\&=&\frac{\alpha^2 \beta^2}{(1+\alpha)(1+\beta)z^2} \int_{0}^{\infty} \left(1+\frac{\alpha}{2}v^2 \right)\left(1+\frac{\beta}{2z^2}v^2 \right)e^{-v\left(\alpha+\frac{\beta}{z} \right)}vdv \\&=&\frac{K}{z^2} \int_{0}^{\infty} \left(1+\frac{v^2}{2}\left(\alpha+\frac{\beta}{z^2} \right)+\frac{\alpha \beta}{4z^2}v^4 \right)e^{-v\left(\alpha+\frac{\beta}{z} \right)}vdv 
\\&=&\frac{K}{z^2} \left\lbrace\int_{0}^{\infty} e^{-v\left(\alpha+\frac{\beta}{z} \right)}vdv + \frac{\alpha+\frac{\beta}{z^2}}{2} \int_{0}^{\infty} e^{-v\left(\alpha+\frac{\beta}{z} \right)}v^3dv + \frac{\alpha \beta}{4z^2} \int_{0}^{\infty} e^{-v\left(\alpha+\frac{\beta}{z} \right)}v^5dv \right\rbrace \\&=&\frac{K}{z^2} \left\lbrace \phi(z)^2 \int_{0}^{\infty} te^{-t}dt+\frac{1}{2}\left(\alpha+\frac{\beta}{z^2} \right)\phi(z)^4 \int_{0}^{\infty} t^3e^{-t}dt+\frac{\alpha \beta}{4z^2}\phi(z)^6 \int_{0}^{\infty} t^5e^{-t}dt \right\rbrace,
\end{eqnarray*}
where, $K=\frac{\alpha^2 \beta^2}{(1+\alpha)(1+\beta)}$ and $\phi(z)=\frac{z}{\alpha z+\beta}$. The last expression is obtained by substituting $v\left(\alpha+\frac{\beta}{z} \right)=t$. On simplification, we obtain (\ref{eqpdf3}).
\end{proof}
The possible shapes of the pdf (\ref{eqpdf3}) of the ratio $Z$ for $\alpha=0.8$ and different values of $\beta$, and for $\beta=0.8$ and different values of $\alpha$, can be provided in Figures. It is observed that as $\alpha$ increases the plot becomes steeper, and as $\beta$ increases the plot becomes flatter. The effects of the parameters are evident from these graphs. In view of these graphs, the proposed distribution appears to be unimodal and right skewed.

\section{Moments and related measures}
\label{sec:3}
In this section, the moments for the random variable $Z$ is obtained. It is observed that only fractional moments of $Z$ exists. 
\begin{thm} \label{Th3}
If $Z$ is a random variable with pdf given by (\ref{eqpdf3}), then its $k$-th moment can be expressed as
\begin{align} \label{eqmoment1}
E\left(Z^k \right)= \left(\frac{1}{\alpha +1}\right)\left(\frac{1}{\beta +1}\right)\left(\frac{\beta}{\alpha} \right)^k \Biggl[ B(k+1,1-k)+3\beta B(k+3,1-k) \\ \nonumber
+3\alpha B(k+1,3-k)+30 B(k+3,3-k) \Biggr], 	
\end{align}
where $-1<k<1$ and $B(.)$ denotes Beta function (or Euler's function of the first kind). 
\end{thm}

\begin{proof}
Using (\ref{eqpdf3}) we obtain,
\begin{align} \label{eq1}
E\left(Z^k \right)= \int_{0}^{\infty} z^k f(z)dz = C \Biggl[ \int_{0}^{\infty} \frac{z^k}{(\alpha z+\beta)^2}dz+ 3\alpha \int_{0}^{\infty} \frac{z^{k+2}}{(\alpha z+\beta)^4}dz \\ \nonumber
+ 3\beta \int_{0}^{\infty} \frac{z^k}{(\alpha z+\beta)^4}dz+ 30\alpha \beta \int_{0}^{\infty} \frac{z^{k+2}}{(\alpha z+\beta)^6}dz \Biggr],
\end{align}
where $C=\left(\frac{\alpha^2}{1+\alpha} \right)\left(\frac{\beta^2}{1+\beta} \right)$. \\
Using the transformation $t=\left(\frac{\alpha}{\beta} \right)z$ and Beta function $B(p,q)=\int_{0}^{\infty} \frac{t^{p-1}}{(1+t)^{p+q}}dt$, we obtain
\begin{align} \label{eq2}
\int_{0}^{\infty} \frac{z^k}{(\alpha z+\beta)^2}dz= \frac{1}{\beta^2} \left(\frac{\beta}{\alpha} \right)^{k+1} B(k+1,1-k)
\end{align}
\begin{align} \label{eq3}
\int_{0}^{\infty} \frac{z^{k+2}}{(\alpha z+\beta)^4}dz= \frac{1}{\beta^4} \left(\frac{\beta}{\alpha} \right)^{k+3} B(k+3,1-k)
\end{align}
\begin{align} \label{eq4}
\int_{0}^{\infty} \frac{z^k}{(\alpha z+\beta)^4}dz= \frac{1}{\beta^4} \left(\frac{\beta}{\alpha} \right)^{k+1} B(k+1,3-k)
\end{align}
\begin{align} \label{eq5}
\int_{0}^{\infty} \frac{z^{k+2}}{(\alpha z+\beta)^6}dz= \frac{1}{\beta^6} \left(\frac{\beta}{\alpha} \right)^{k+3} B(k+3,3-k)
\end{align}
Using equations (\ref{eq2}), (\ref{eq3}), (\ref{eq4}), (\ref{eq5}), and upon simplification, from equation (\ref{eq1}), the Theorem (\ref{Th3}) follows provided $-1<k<1$.
\end{proof}
It is evident from Theorem (\ref{Th3}) that only the fractional moments of order $-1<k<1$ of $Z$ exist. The plots for the fractional moments for $\alpha=0.8$ and different values of $\beta$, and for $\beta=0.8$ and different values of $\alpha$, can be shown using figures, when $-0.5 \leq k \leq 0.5$.

\section{Entropy of $Z=|\frac{X}{Y}|$}
\label{sec:4}
The entropy of a continuous random variable $Z$ is a measure of variation of uncertainty and has applications in many fields such as physics, engineering, and economics, among others. \\
For a non-negative continuous random variable $Z$ with pdf $f(z)$, Shannon measure of entropy is defined as 
\begin{align*}
H(f)=E\left[- \ln f(z) \right]= - \int_{0}^{\infty} \left(\ln f(z) \right)f(z)dz
\end{align*}
Using the expression for the pdf (\ref{eqpdf3}) in the above equation and simplifying, we obtain
\begin{align} \label{ent1}
E\left[- \ln f_Z(z) \right]= \ln \left[\frac{(\alpha+1)(\beta+1)}{\alpha^2 \beta^2} \right]- E\left[g(Z) \right] , 
\end{align}
where $g(z)=\ln \left[\frac{1}{(\alpha z+\beta)^2}+\frac{3\left(\alpha z^2+\beta \right)}{(\alpha z+\beta)^4}+\frac{30\alpha \beta z^2}{(\alpha z+\beta)^6} \right]$. \\
For a non-negative continuous random variable $Z$ with pdf $f(z)$, Rényi entropy is defined as
\begin{align*}
H_R(\gamma)=\frac{1}{1-\gamma} \ln \left[\int_{0}^{\infty} f^\gamma (z)dz \right]
\end{align*} 
for $\gamma >0, \gamma \neq 1$. \\
Using the expression for the pdf (\ref{eqpdf3}) in the above equation and simplifying, we obtain
\begin{align} \label{ent2}
H_R(\gamma)=\frac{\gamma}{1-\gamma} \ln \left[\frac{\alpha^2 \beta^2}{(\alpha +1)(\beta +1)} \right]+ \frac{1}{1-\gamma} \ln \left[\int_{0}^{\infty} h^\gamma (z)dz \right]
\end{align}
In physics, Tsallis entropy is a generalization of the standard Boltzmann-Gibbs entropy. For an absolutely continuous non-negative random variable $Z$ with pdf $f(z)$, Tsallis entropy (also called q-entropy) is defined as
\begin{align*}
S_q(Z)=\frac{1}{1-q} \ln \left[1-\int_{0}^{\infty} f^q (z)dz \right],
\end{align*}
for $q >0, q \neq 1$. \\
Using the expression for the pdf (\ref{eqpdf3}) in the above equation and simplifying, we obtain
\begin{align} \label{ent3}
S_q(Z)=\frac{1}{1-q} \ln \left[1- \left(\frac{\alpha^2 \beta^2}{(\alpha +1)(\beta +1)} \right)^q \int_{0}^{\infty} h^q (z)dz \right],
\end{align}
where $h(z)=\frac{1}{(\alpha z+\beta)^2}+\frac{3\left(\alpha z^2+\beta \right)}{(\alpha z+\beta)^4}+\frac{30\alpha \beta z^2}{(\alpha z+\beta)^6}$. \\
It can be observed that none of the entropies in equations (\ref{ent1}), (\ref{ent2}), and (\ref{ent3}) can be evaluated analytically in closed form and hence require some appropriate numerical computations. 

\section{Characterization}
The characterization relation is a certain distributional or statistical property of a statistic or statistics that uniquely determines the associated probability distribution. In order to apply a particular probability distribution to some real world data, it is necessary to characterize it first subject to certain conditions. There are various methods of characterizations to identify the distribution of a real data set. One of the most important method of characterization is the method of truncated
moments. For more details on characterization of distributions, one may refer Kagan et el. (1973), Galambos and Kotz (1978), Nagaraja (2006), Ahsanullah (2017) etc. We shall prove the characterization theorems by the right (left) truncated conditional expectations of $Z^k$ for $-1<k<1$, by considering a product of reverse hazard rate (hazard rate) and another function of the truncated point. \\
We shall need the following assumption and lemmas to prove the characterization results.
\begin{assmp} \label{Assmp1}
Suppose the random variable $Z$ is having absolutely continuous cdf $F(z)$ and pdf $f(z)$. We assume that $\inf \left\lbrace z| F(z)>0 \right\rbrace =0,$ and $\sup \left\lbrace z| F(z)<1 \right\rbrace =\infty$. We also assume that $f(z)$ is differentiable for all $z$ and $E\left(Z^k \right)$ exists for all $-1<k<1$.
\end{assmp}
\begin{lem} \label{Lem1}
Suppose a non-negative random variable $Z$ satisfies assumption \ref{Assmp1}. Then, if the k-th order right truncated moment $E\left(Z^k | Z \leq z \right)=g(z)r(z)$, where $r(z)=\frac{f(z)}{F(z)}$ is the reverse hazard rate associated with the distribution of $Z$, and $g'(z)$ exists for all $z>0$, we have $f(z)=c e^{\int_{0}^{z} \frac{u^k - g'(u)}{g(u)}du}$, where $c$ is a constant and $c=\int_{0}^{\infty}f(z)dz =1$.
\end{lem}
\begin{proof}
$ E\left(Z^k | Z \leq z \right)=\frac{\int_{0}^{z}u^k f(u)du}{F(z)} \Rightarrow \int_{0}^{z}u^k f(u)du=g(z)f(z) $. Differentiating both sides with respect to $z$, we obtain, $ z^k f(z)=g'(z)f(z)+g(z)f'(z) $, or, $ \frac{f'(z)}{f(z)}=\frac{z^k -g'(z)}{g(z)} $ and upon integrating both sides we get $f(z)=c e^{\int_{0}^{z} \frac{u^k - g'(u)}{g(u)}du}$, where $ c=\int_{0}^{\infty}f(z)dz =1 $.
\end{proof}
\begin{lem} \label{Lem2}
Suppose a non-negative random variable $Z$ satisfies assumption \ref{Assmp1}. Then, if the k-th order left truncated moment $E\left(Z^k | Z \geq z \right)=g_1(z)h(z)$, where $h(z)=\frac{f(z)}{1-F(z)}$ is the hazard rate associated with the distribution of $Z$, and ${g_1}'(z)$ exists for all $z>0$ with the condition that $ \int_{z}^{\infty} \frac{u^k+{g_1}'(u)}{{g_1}(u)}du <0 $ for $z>0$, then $f(z)=c e^{-\int_{z}^{\infty} \frac{u^k+{g_1}'(u)}{{g_1}(u)}du}$, where $c$ is a constant and $c=\int_{0}^{\infty}f(z)dz =1$.
\end{lem}
\begin{proof}
$ E\left(Z^k | Z \geq z \right)=\frac{\int_{0}^{z}u^k f(u)du}{1-F(z)} \Rightarrow \int_{z}^{\infty}u^k f(u)du=g_1(z)f(z) \Rightarrow E\left(Z^k \right)-\int_{0}^{z}u^k f(u)du=g_1(z)f(z) $. Differentiating and upon simplification, we get $ \frac{f'(z)}{f(z)}=-\frac{z^k+{g_1}'(z)}{{g_1}(z)} $ and upon integrating the proof follows. 
\end{proof}
\begin{thm} \label{Th4}
Let a non-negative random variable $Z$ satisfies assumption \ref{Assmp1}. Then \\
$ E\left(Z^k | Z \leq z \right)=g(z)r(z) $, where $r(z)=\frac{f(z)}{F(z)}$, and $g(z)=\frac{I_k(z)}{\left(\frac{\alpha^2}{1+\alpha} \right)\left(\frac{\beta^2}{1+\beta} \right)\left[\frac{1}{(\alpha z+\beta)^2}+\frac{3\left(\alpha z^2+\beta \right)}{(\alpha z+\beta)^4}+\frac{30\alpha \beta z^2}{(\alpha z+\beta)^6} \right]}$ iff $Z$ has the pdf $ f(z)=\left(\frac{\alpha^2}{1+\alpha} \right)\left(\frac{\beta^2}{1+\beta} \right)\left[\frac{1}{(\alpha z+\beta)^2}+\frac{3\left(\alpha z^2+\beta \right)}{(\alpha z+\beta)^4}+\frac{30\alpha \beta z^2}{(\alpha z+\beta)^6} \right] $, where $ I_k(z)=\int_{0}^{z}u^k f(u)du $ denotes the k-th incomplete moment of $Z$.
\end{thm}
\begin{proof}
Suppose that $ E\left(Z^k | Z \leq z \right)=g(z)\frac{f(z)}{F(z)}$ where the random variable $Z$ is having pdf (\ref{eqpdf3}). Therefore, $ E\left(Z^k | Z \leq z \right)=\frac{\int_{0}^{z}u^k f(u)du}{F(z)}=\frac{g(z)f(z)}{F(z)} \Rightarrow g(z)=\frac{\int_{0}^{z}u^k f(u)du}{f(z)} $, proving necessity. \\
To prove sufficiency, let 
\begin{align} \label{eq6}
g(z)=\frac{I_k(z)}{\left(\frac{\alpha^2}{1+\alpha} \right)\left(\frac{\beta^2}{1+\beta} \right)\left[\frac{1}{(\alpha z+\beta)^2}+\frac{3\left(\alpha z^2+\beta \right)}{(\alpha z+\beta)^4}+\frac{30\alpha \beta z^2}{(\alpha z+\beta)^6} \right]} 
\end{align}
where $ I_k(z)=\int_{0}^{z}u^k f(u)du $. \\
Now, from Lemma (\ref{Lem1}), we have, 
\begin{align} \label{eq7}
f(z)=e^{\int_{0}^{z} \frac{u^k - g'(u)}{g(u)}du} \Rightarrow \frac{z^k - g'(z)}{g(z)}=\frac{f'(z)}{f(z)} 
\end{align}
Using (\ref{eq6}) and upon simplification, we get
\begin{align} \label{eq8}
\frac{z^k\frac{f(z)}{\left(\frac{\alpha^2}{1+\alpha} \right)\left(\frac{\beta^2}{1+\beta} \right)\left[\frac{1}{(\alpha z+\beta)^2}+\frac{3\left(\alpha z^2+\beta \right)}{(\alpha z+\beta)^4}+\frac{30\alpha \beta z^2}{(\alpha z+\beta)^6} \right]}-g'(z)}{g(z)}=\frac{\left[-\frac{2\alpha}{(\alpha z+\beta)^3}+\frac{6(-\alpha^2 z^2+\alpha \beta z -2\alpha \beta)}{(\alpha z+\beta)^5}+\frac{60\alpha \beta (-2\alpha z^2 + \beta z)}{(\alpha z+\beta)^7} \right]}{\left[\frac{1}{(\alpha z+\beta)^2}+\frac{3\left(\alpha z^2+\beta \right)}{(\alpha z+\beta)^4}+\frac{30\alpha \beta z^2}{(\alpha z+\beta)^6} \right]}
\end{align} 
Combining (\ref{eq7}) and (\ref{eq8}) we get $ f(z)=\left(\frac{\alpha^2}{1+\alpha} \right)\left(\frac{\beta^2}{1+\beta} \right)\left[\frac{1}{(\alpha z+\beta)^2}+\frac{3\left(\alpha z^2+\beta \right)}{(\alpha z+\beta)^4}+\frac{30\alpha \beta z^2}{(\alpha z+\beta)^6} \right] $.
\end{proof}
\begin{thm} \label{Th5}
Let a non-negative random variable $Z$ satisfies assumption \ref{Assmp1}. Then $ E\left(Z^k | Z \geq z \right)=\frac{g_1(z)f(z)}{1-F(z)} $, where $g_1(z)= \frac{E\left(Z^k \right)-g(z)f(z)}{f(z)}$, $E\left(Z^k \right)$ is given by (\ref{eqmoment1}) for $-1<k<1$, and $g(z)$ is given by (\ref{eq6}), iff $Z$ has the pdf $ f(z)=\left(\frac{\alpha^2}{1+\alpha} \right)\left(\frac{\beta^2}{1+\beta} \right)\left[\frac{1}{(\alpha z+\beta)^2}+\frac{3\left(\alpha z^2+\beta \right)}{(\alpha z+\beta)^4}+\frac{30\alpha \beta z^2}{(\alpha z+\beta)^6} \right] $.
\end{thm}
\begin{proof}
Proceeding in the same way as in Theorem \ref{Th4}, following similar arguments, the proof of Theorem \ref{Th5} easily follows using Lemma \ref{Lem2}.
\end{proof}

%\clearpage


\begin{thebibliography}{99}

\bibitem{} Abulebda, M., Pathak, A. K., Pandey, A., and Tyagi, S. (2022). On a Bivariate XGamma Distribution Derived from Copula. Statistica, 82(1), 15-40.

\bibitem{} Ali, M. M., Pal, M., and Woo, J. (2007). On the ratio of inverted gamma variates. Austrian Journal of Statistics, 36(2), 153-159.

\bibitem{} Alizadeh, M., Afshari, M., Ranjbar, V., Merovci, F., and Yousof, H. M. (2023). A novel XGamma extension: applications and actuarial risk analysis under the reinsurance data. São Paulo Journal of Mathematical Sciences, 1-31.

\bibitem{} Bowman, K. O., Shenton, L. R., and Gailey, P. C. (1998). Distribution of the ratio of gamma variates. Communications in Statistics-Simulation and Computation, 27(1), 1-19.

\bibitem{} Cordeiro, G. M., Altun, E., Korkmaz, M. C., Pescim, R. R., Afify, A. Z., and Yousof, H. M. (2020). The xgamma family: Censored regression modelling and applications. Revstat-Statistical Journal, 18(5), 593-612.

\bibitem{} Cox, D. R. and Hinkley, D. V. (1979). Theoretical statistics, CRC Press.

\bibitem{} Nadarajah, S. (2010). Distribution properties and estimation of the ratio of independent Weibull random variables. AStA Advances in Statistical Analysis, 94, 231-246.

\bibitem{} Pathak, A., Kumar, M., Singh, S. K., Singh, U., Tiwari, M. K., and Kumar, S. (2023). E-Bayesian inference for xgamma distribution under progressive type II censoring with binomial removals and their applications. International Journal of Modelling and Simulation, 1-20.

\bibitem{} Sackey, E. K. and Smith, G. D. (2009). Empirical distribution models for slenderness and aspect ratios of core particles of particulate wood composites. Wood and Fiber Science, 255-266.

\bibitem{} Sen, S., Afify, A. Z., Al-Mofleh, H., and Ahsanullah, M. (2019). The quasi xgamma-geometric distribution with application in medicine. Filomat, 33(16), 5291-5330.

\bibitem{} Sen, S. and Chandra, N. (2017). The quasi xgamma distribution with application in bladder cancer data. Journal of data science, 15(1), 61-76.

\bibitem{} Sen, S., Chandra, N., and Maiti, S. S. (2018). Survival estimation in xgamma distribution under progressively type-II right censored scheme. Model Assisted Statistics and Applications, 13(2), 107-121.

\bibitem{} Sen, S., Maiti, S. S., and Chandra, N. (2016). The xgamma distribution: statistical properties and application. Journal of Modern Applied Statistical Methods, 15(1), 774-788.

\bibitem{} Worzyk, T. B., Bergkvist, M., Nordberg, P., and Tornkvist, C. (1997). Breakdown voltage of polypropylene laminated paper (PPLP) in plain samples and a full scale cable. In IEEE 1997 Annual Report Conference on Electrical Insulation and Dielectric Phenomena, Vol. 1, 329-333.

\bibitem{} Yadav, A. S., Maiti, S. S., and Saha, M. (2021). The inverse xgamma distribution: statistical properties and different methods of estimation. Annals of Data Science, 8, 275-293.

\end{thebibliography}
\end{document}